\documentclass[runningheads]{llncs}
\usepackage{graphicx}

\usepackage{tikz}
\usepackage{comment}
\usepackage{amsmath,amssymb} 
\usepackage{color}

\usepackage{multirow}
\usepackage{multicol}
\usepackage{subcaption}
\usepackage{caption} 
\usepackage{algorithm}
\DeclareMathOperator*{\argminB}{argmin}

\usepackage[accsupp]{axessibility}

\usepackage{xspace}
\makeatletter
\DeclareRobustCommand\onedot{\futurelet\@let@token\@onedot}
\def\@onedot{\ifx\@let@token.\else.\null\fi\xspace}

\def\eg{\emph{e.g}\onedot} 
\def\ie{\emph{i.e}\onedot}

\def\etal{\emph{et al}\onedot}
\makeatother

\usepackage[capitalize]{cleveref}
\crefname{section}{Sec.}{Secs.}
\Crefname{section}{Section}{Sections}
\Crefname{table}{Table}{Tables}

\usepackage[width=122mm,left=12mm,paperwidth=146mm,height=193mm,top=12mm,paperheight=217mm]{geometry}

\begin{document}
\pagestyle{headings}
\mainmatter
\def\ECCVSubNumber{1586}  

\title{Perception-Distortion Balanced ADMM \\ 
Optimization for Single-Image Super-Resolution} 

\titlerunning{PD-Balanced ADMM Optimization for SISR}
%
\author{Yuehan Zhang\inst{1}\orcidID{0000-0002-5017-0097}
\and Bo Ji\inst{1}\orcidID{0000-0002-5718-2074} 
\and Jia Hao\inst{2}\orcidID{0000-0002-0909-9444}
\and Angela Yao\inst{1}\orcidID{0000-0001-7418-6141}}
\authorrunning{Y. Zhang et al.}
%
\institute{National University of Singapore, \and HiSilicon Technologies, Shanghai\\\email{\{zyuehan,jibo,ayao\}@comp.nus.edu.sg, hao.jia@huawei.com} }


\maketitle

\begin{abstract}In image super-resolution, both pixel-wise accuracy and perceptual fidelity are desirable. However, most deep learning methods only achieve high performance in one aspect due to the perception-distortion trade-off, and works that successfully balance the trade-off rely on fusing results from separately trained models with ad-hoc post-processing. In this paper, we propose a novel super-resolution model with a low-frequency constraint (LFc-SR), which balances the objective and perceptual quality through a single model and yields super-resolved images with high PSNR and perceptual scores. We further introduce an ADMM-based alternating optimization method for the non-trivial learning of the constrained model. Experiments showed that our method, without cumbersome post-processing procedures, achieved the state-of-the-art performance. The code is available at \url{https://github.com/Yuehan717/PDASR}.

\end{abstract}

\section{Introduction}
\label{sec:intro}
\begin{figure}[h]
    \begin{minipage}[h]{0.50\linewidth}
    \centering
    \subfloat[Ground Truth]{
        \includegraphics[width=0.45\linewidth]{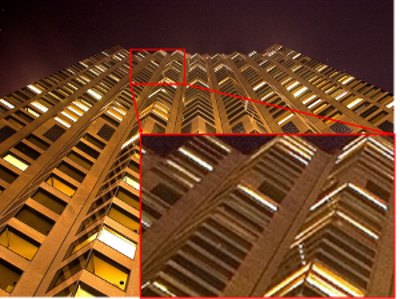}
    }
    \subfloat[Objective-aimed]{
        \includegraphics[width=0.45\linewidth]{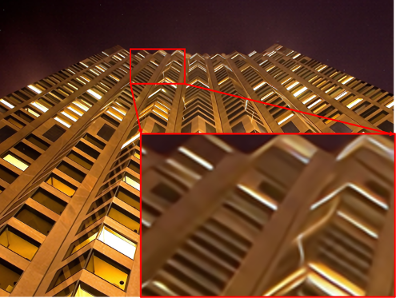}
    }
    
    \subfloat[Perceptual-aimed]{
        \includegraphics[width=0.45\linewidth]{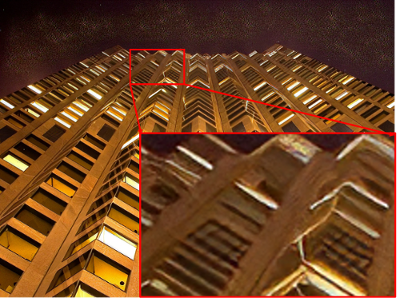}
    }
    \subfloat[Balanced Quality]{
        \includegraphics[width=0.45\linewidth]{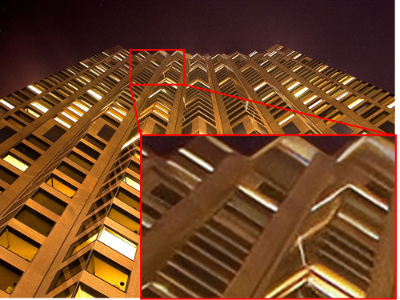}
    }
    \end{minipage}
    \medskip
    \begin{minipage}[h]{0.48\linewidth}
    \centering
    \subfloat[SOTAs performance]{
    \includegraphics[width=0.95\linewidth]{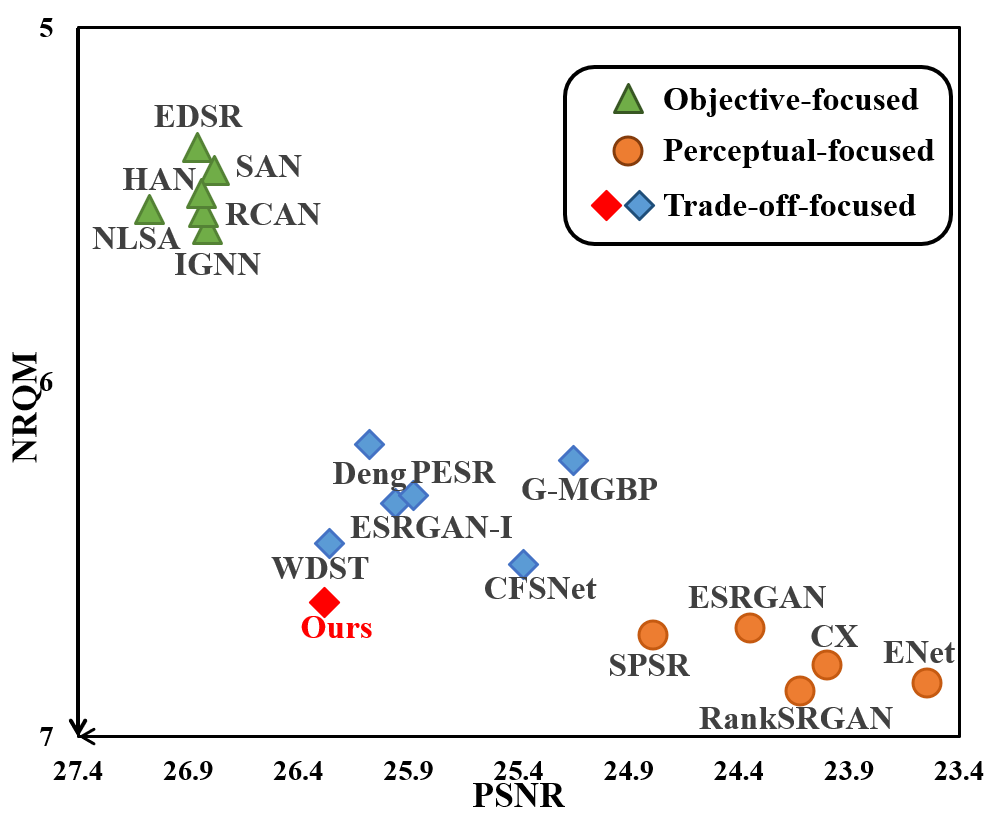}
    }
    \end{minipage}
    \caption{(a)-(d) are ground truth from Urban100\cite{Huang-CVPR-2015} and $\times4$ super-resolved images. 
    Note that (b) produced by NLSA~\cite{mei2021image} is blurry in detailing, while (c) from RankSRGAN~\cite{zhang2019ranksrgan} has unnatural patterns; our balanced approach in (d) mitigates both types of artifacts. 
    (e) is performance of state-of-the-art $\times 4$ SISR models on Urban100. Higher PSNR and NRQM scores indicate better objective and perceptual quality.
    Our method reaches state-of-the-art while being single-shot.}
    \label{fig:compare}
\end{figure}

Single image super-resolution (SISR) recovers a high-resolution (HR) image from a low-resolution (LR) input. There are two types of quality assessments for super-resolved images: objective quality, evaluated by PSNR and SSIM, and perceptual quality, based on metrics such as NRQM~\cite{ma2017learning} and LPIPS~\cite{zhang2018unreasonable}.
SISR methods aiming at high objective quality~\cite{dai2019second,kim2016accurate,kim2016deeply,lim2017enhanced,mei2021image,niu2020single,zhang2018image,zhang2018residual} achieve high PSNR values, but the results look blurry. Another line of research focuses on improving perceptual quality~\cite{ledig2017photo,ma2020structure,mechrez2018maintaining,park2018srfeat,wang2018esrgan,zhang2019ranksrgan}, which produces HR images that are visually shaper, but have lower PSNR scores and unrealistic patterns (see comparisons in~\cref{fig:compare}).

An ideally reconstructed image is similar to the ground truth HR image, with limited distortion \emph{and} high perceptual quality, where distortion refers to a drop in objective quality. However, most deep learning methods can achieve high performance in only one of the two qualities. This perception-distortion (PD) trade-off is rooted in the supervised training process of SISR methods. Since perception and distortion measurements are incoherent with each other, optimization in one direction naturally leads to sacrifice in the other~\cite{blau2018perception,freirich2021theory}. 
A recently emerging line of work seeks to balance or improve both the objective and perceptual quality of SISR images \cite{deng2018enhancing,deng2019wavelet,vu2018perception,vasu2018analyzing,wang2018esrgan,navarrete2018multi,shoshan2019dynamic,lee2020smoother,wang2019cfsnet}. Some researchers have studied the trade-off problem through model optimization~\cite{vasu2018analyzing,navarrete2018multi,shoshan2019dynamic,lee2020smoother,wang2019cfsnet}. However, they have only focused on traversing the PD trade-off instead of improving both the objective and perceptual quality. Others bypassed the optimization incoherence and merged the models or model outputs trained separately for the two objectives via post-processing~\cite{deng2018enhancing,deng2019wavelet,vu2018perception,wang2018esrgan}. Some methods~\cite{deng2018enhancing,deng2019wavelet} achieved excellent results, but their post-processing can be computationally costly.

Previous studies have aligned image quality with different image frequency subbands. Low-frequency (LF) information captures the overall scene structure~\cite{ramamonjisoa2021single,zhou2020guided}, while high-frequency (HF) details are critical to achieving high perceptual quality~\cite{ledig2017photo,zhou2020guided}. This observation has been leveraged to reconstruct details in single-image SR~\cite{singh2020wdn,zhou2020guided} and suppress misalignment in real-world video SR~\cite{yang2021real}. However, it has not been used to explicitly balance the objective and perceptual quality in model optimization.

This paper present a new low-frequency constrained SISR model (LFc-SR) model that improves both objective and perceptual quality. The model has two stages: stage one focuses on super-resolution for objective quality, while stage two subsequently refines the image for perceptual quality (see~\cref{fig:arch}). At our model's core is a similarity constraint that keeps the low-frequency subbands of both stages similar during optimization. The constraint allows the optimization of perceptual quality to be oriented towards the high-frequency bands while ensuring that the overall scene retains the effects of objective-focused learning. As such, our method yields images with high performance on both qualities in a one-shot manner.

Our key novelty is in our formulation of LFc-SR training as constrained multi-objective optimization. Enforcing constraints within deep learning is non-trivial with challenges to find satisfying solutions. To cooperate the low-frequency constraint, we designed a novel ADMM-based alternating optimization method. The alternating direction method of multipliers (ADMM) was originally introduced as a tool for convex optimization problems, decomposing the constrained optimization into solvable substeps~\cite{boyd2011distributed}. However, it has also been widely applied to non-convex problems, including deep neural network training~\cite{kiaee2016alternating,zhao2018admm,li2017deep,zhang2018systematic,takapoui2020simple,leng2018extremely}. While finding the optimum for non-convex problems is not guaranteed, ADMM often finds satisfying approximations. We refer to our optimization method as PD-ADMM, as it aims at mitigating the perception-distortion (PD) trade-off.

To summarize our contributions in this paper:
\begin{enumerate}
    \item We propose LFc-SR, a novel SISR model trained with a low-frequency constraint, to balance the perception-distortion trade-off.
    \item We propose a novel formulation of SISR as a constrained multi-objective optimization problem and show how to incorporate an ADMM-based method (PD-ADMM) for solutions that balance the perception-distortion trade-off. 
    \item Our LFc-SR model learned with the PD-ADMM optimization scheme yields HR images with high objective and perceptual quality. The proposed one-shot model requires significantly less computational expense than competing trade-off methods that rely on heavy post-processing.
\end{enumerate}

\section{Related Work}
\label{sec:related-work}
\subsection{Objective Quality versus Perceptual Quality}
SISR methods designed for objective quality are deep neural networks that use pixel-wise losses, such as $L_1$ and MSE. Starting with the CNN-based SRCNN~\cite{dong2015image}, later variants leverage residual~\cite{kim2016accurate,lim2017enhanced,ledig2017photo}, dense~\cite{zhang2018residual,tong2017image}, and attention mechanisms~\cite{mei2021image,niu2020single,dai2019second}. These works strongly focus on architecture design and achieve high PSNR and SSIM scores. However, their results are blurry and evaluated poorly on perceptual metrics, especially at larger scaling factors like $\times4$.

Many SISR methods often incorporate GANs~\cite{ledig2017photo,wang2018esrgan,park2018srfeat,mechrez2018maintaining,zhang2019ranksrgan,ma2020structure} and combine the adversarial loss with a content loss to improve the perceptual quality. Common content losses include the VGG~\cite{ledig2017photo} and contextual loss~\cite{mechrez2018maintaining}. GAN-based models yield sharper lines and more high-frequency details. They achieve high scores on perceptual measures such as LPIPS~\cite{zhang2018unreasonable} and NRQM~\cite{ma2017learning}, attributed to implicit distribution learning. At the same time, these methods suffer from unrealistic artifacts resulting from adversarial training.

\subsection{Perception-Distortion Trade-off}
Most state-of-the-art methods perform well either in objective or perceptual quality, but not in both. This trade-off phenomenon was explored by Blau~\etal~\cite{blau2018perception}; they attributed it to the incoherence between distortion and perception measurements. Some methods have tried to control the compromise between the two qualities. 
One solution was to fuse either two models~\cite{wang2018esrgan} or model outputs~\cite{vu2018perception,deng2019wavelet,deng2018enhancing} trained separately for objective and perceptual quality. The most notable of these post-processing works is WDST~\cite{deng2019wavelet}, which, like our work, set out to take separate frequency band considerations. A significant difference between our work and WDST is that WDST merges wavelet channels from the outputs of two separately trained networks, whereas our method is one-shot with a single model. Moreover, as WDST relies on a style-transfer~\cite{gatys2016image} for the fusion, their inference procedure is computationally expensive and time-consuming, as the output channels are iteratively merged at test time. Our one-shot inference consumes significantly less computational resources and is much more efficient.

In addition to post-processing, another line of balancing approaches focuses on the training strategy. They balance the loss terms~\cite{vasu2018analyzing} or introduce a controlling factor as training inputs~\cite{navarrete2018multi,shoshan2019dynamic,wang2019cfsnet,lee2020smoother}. For example, Vasu~\etal~\cite{vasu2018analyzing} directly added MSE, VGG, and adversarial loss together and weighted them in different proportions. CFSNet~\cite{wang2019cfsnet} achieved a test-time transition between perceptual and objective quality by using an extra input factor $\alpha$ in their training strategy. These approaches create a smooth transition between the two qualities; but do not attempt to improve both of them. In contrast, our method focuses on the training strategy \emph{and} improves both qualities.
\section{Method}
\label{sec:method}
\subsection{Revisiting the Multi-Objective SR Formulation}
\label{sec:formulation}
SISR recovers a high-resolution image $Y \in \mathbb{R}^{H\times W}$, given an input low-resolution image $X \in \mathbb{R}^{h \times w}$ via a neural network $G$, \ie $Y = G(X)$. Here, $h \times w$ and $H\times W$ are input and output image heights and widths respectively. As it is challenging to gather real-world pairs of images in low- and high-resolution, $X$ is typically generated from a ground truth high-resolution image $\hat{Y}$ via a downsampling function, \ie $X = f_{\downarrow}(\hat{Y})$. Bicubic downsampling is a commonly used $f_{\downarrow}(\cdot)$, with standard downsampling factors of $\times2$ or $\times4$, \eg $H = 2\times h, W = 2\times w$. 

Considering both distortion and perception, a simple way to learn $G(\cdot)$ is by weighting different loss terms $L_O$ and $L_P$ with weights $\lambda_O$ and $\lambda_P$\footnote{Throughout the paper, we will use  $O$ and $P$ (either subscript or superscript) to denote objective- and perception-focused items respectively.} for objective and perceptual quality respectively, \ie
\begin{equation}\label{eq:naivesum}
    L_G = \lambda_O \cdot L_{O}(Y) + \lambda_{P} \cdot L_{P}(Y).
\end{equation}
\noindent{However, this does not achieve a good PD trade-off because of the incoherence between perception and distortion losses~\cite{blau2018perception}.
Specifically, minimizing $L_O$ (\ie~with MSE, $L_1$) leads to low expectation of pixel-wise error, whereas $L_P$ (\ie~adversarial loss combined with contextual loss) targets distribution-based learning for sharp results~\cite{blau2018perception}}. It is therefore common for methods seeking a good trade-off to learn two separate networks and fuse the results via post-processing, \ie 
\begin{equation}
\begin{aligned}
    Y = F(Y^O, Y^P), \quad \text{where} \quad Y^O = G_O(X) \text{, and}\; Y^P = G_P(X),
\end{aligned}
\label{eq:image_interpolation}
\end{equation}
where $G_O$ and $G_P$ are two separately trained models 
and $F$ is some fusion module. $F$ can be quite complex and computationally expensive. For example, in Deng~\cite{deng2018enhancing} and WDST~\cite{deng2019wavelet}, $F$ is a style transfer module applied during inference. 

Inspired by previous research in human perception~\cite{kim2017deep,zhou2020guided} and the direct link between low- and high-frequency information to objective and perceptual quality~\cite{deng2019wavelet,ledig2017photo}, we take a frequency-banded approach. A na\"ive formulation is to separate the learning objectives for the low- and high-frequency bands, \ie 
\label{sec:arch}
\begin{equation}
    \begin{aligned}\label{eq:freqsep}
        &Y = G(X), \;\;\; \text{where} \;\;\; &L_G = \lambda_O \cdot L_{O}(T_{LF}(Y)) + \lambda_{P} \cdot L_{P}(T_{HF}(Y)).
    \end{aligned}
\end{equation}

\noindent In this formulation, the network $G$ is learned by applying a weighted loss on the low-pass and high-pass decomposed subbands of the output $Y$, based on the respective transformations $T_{LF}$ and $T_{HF}$. Like previous works \cite{deng2019wavelet,zhong2018joint,liu2018multi}, we used the discrete wavelet transform with a Haar wavelet basis for decomposing the image into frequency subbands. Supervising a network based on the loss of~\cref{eq:freqsep} assumes that the different frequency bands can be optimized independently, which is not the case.   
A hard separation increases the difficulty of learning, especially for high-frequency bands $T_{HF}(Y)$, as $L_P$ includes an adversarial loss and is, from our observations, unstable and prone to training collapse.

\subsection{Low-Frequency Constrained SR (LFc-SR)}
\label{sec:LFc-SR}
\begin{figure*}[t!]
    \centering
    \includegraphics[width = 0.95\linewidth]{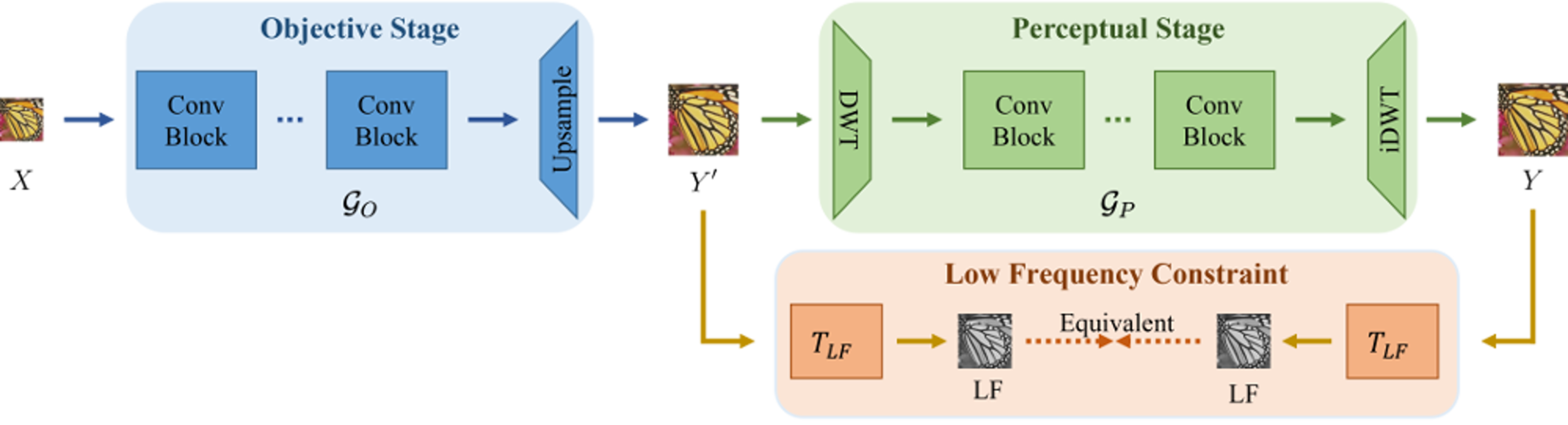}
    \caption{Proposed two-stage model with a low-frequency constraint. The first stage learns high-fidelity LF subbands to yield high objective quality.
    The second stage focuses on the distribution learning of HF subbands to improve perceptual quality for the overall image. Discrete Wavelet Transform (DWT) decreases the spatial size of each channel without information loss, and Inverse Discrete Wavelet Transform (iDWT) recovers the spatial resolution (see \cref{sec:setting} for details). The two stages are trained with the low-frequency constraint to encourage the similarity of the LF subbands of the two outputs.
    }
    \label{fig:arch}
\end{figure*}
Due to the lack of independence, we contend that the optimization of each objective, be it objective or perceptual, must observe the entire band of image frequencies. Our solution to encourage stable learning of high-frequency information of $Y$ while maintaining objective quality is to allow a separate (early stage) output $Y^{\prime}$ for the objective quality. Specifically, we partitioned the model $G$ into two stages, where $G=\{\mathcal{G}_O, \mathcal{G}_P\}$ (see \cref{fig:arch}), \ie 
\begin{equation}
\begin{aligned}
     & Y^{\prime} = \mathcal{G}_O(X), \quad
        Y = \mathcal{G}_P(Y^{\prime}) \;\; 
        \quad s.t.\;\; \mathrm{T}_{LF}(Y) = \mathrm{T}_{LF}(Y^{\prime}),
    \end{aligned}
    \label{eq:stg2_aim}
\end{equation}
where $\mathcal{G}_O(\cdot)$ represents the first objective-focused stage of $G$ and $Y^{\prime}$ is a super-resolved image with high objective quality. 
$\mathcal{G}_P(\cdot)$ is the perception-focused stage targeting a further improvement on high-frequency detailing that takes $Y^{\prime}$ as input. To encourage the final output $Y$ to maintain high objective quality, we placed a constraint that the low-frequency subbands of $Y$ should be equal to those of $Y^{\prime}$, which in practice encourages the two to be similar. 

The two stages accordingly have different loss functions. Outputs from $\mathcal{G}_O$ are supervised by a $L_1$ loss:
\begin{equation}
        L_O = L_1(Y^\prime,\hat{Y})
    \label{eq:stage1}
\end{equation}
where $\hat{Y}$ is the ground truth for the training input $X$. 
For $\mathcal{G}_P$, we used the contextual loss~\cite{mechrez2018maintaining} together with adversarial training for high perceptual quality. The contextual loss approximates KL-divergence between the deep features of super-resolved and ground-truth images, and we refer to Mechrez \etal \cite{mechrez2018contextual,mechrez2018maintaining} for further details. The overall loss function for $\mathcal{G}_P$ is as follows:
\begin{equation}
    \begin{aligned}
        L_P = 
        \lambda_{CX} \cdot L_{CX}(Y,\hat{Y}) + \lambda_{D} \cdot L_1(f_{\downarrow}(Y),f_{\downarrow}(\hat{Y})) + \lambda_{Gen} \cdot L_{Gen}(Y),
    \end{aligned}
    \label{eq:stage2}
\end{equation}
\noindent where $L_{CX}$ is the contextual loss and $f_{\downarrow}(\cdot)$ estimates the low-resolution counterparts of HR images. The $L_1$ loss here is, as introduced in Mechrez~\etal~\cite{mechrez2018contextual}, to boost the spatial structure similarity with the contextual loss, and $L_{Gen}$
represents the generator loss used in adversarial training. 

A straightforward way to implement the equivalence constraint in \cref{eq:stg2_aim} is to apply a $L_1$ regularizer on the outputs of two stages, and the total loss is as following:
\begin{equation}
\begin{aligned}
    \lambda_O \cdot L_O + \lambda_P \cdot L_P + \lambda_r \cdot \left \| \mathrm{T}_{LF}(Y) - \mathrm{T}_{LF}(Y^{\prime}) \right \|_1.
\end{aligned}
\label{eq:ll_reg}
\end{equation}

\noindent Theoretically, regularization will also encourage LF similarity. However, minimizing \cref{eq:ll_reg} requires a difficult-to-achieve balance between $L_O$, $L_P$~\cite{chen2018fast} and the selection of a good $\lambda_r$. To bypass this difficulty, we interpret \cref{eq:stg2_aim} as a multi-objective optimization problem with a combinatorial constraint. ADMM is an established algorithm for solving such constrained optimizations and in the next section, we outline PD-ADMM, our ADMM-based optimization method. 

\subsection{Alternating Optimization with ADMM}
\label{sec:admm}
We designed an alternating training method, PD-ADMM, based on the Alternating Direction Method of Multipliers (ADMM)~\cite{boyd2011distributed} for the optimization of LFc-SR. 

Considering \cref{eq:stg2_aim}, finding solutions for the LFc-SR model is actually a constrained multi-objective optimization problem, \ie there are two different objectives with the constraint on low-frequency coherency:
\begin{equation}
\begin{aligned}
    \min \; L_P + L_O \quad
    s.t.\;\; \space \mathrm{T}_{LF}(Y) = \mathrm{T}_{LF}(Y^{\prime}),
\end{aligned}
\label{eq:formu_constrained}
\end{equation}
where $L_P$ and $L_O$ are loss functions in \cref{sec:LFc-SR}, and the constraint is the same as that in \cref{eq:stg2_aim}.
Since the model parameters are variables in this optimization, we express \cref{eq:formu_constrained} as functions of $\theta_P$ and $\theta_O$, the parameters of $\mathcal{G}_P$ and $\mathcal{G}_O$ :
\begin{equation}
\begin{aligned} 
\min_{\theta_P,\theta_O}\;P(\theta_P) + O(\theta_O) 
\quad s.t. \;\; \space \mathrm{LF_p}(\theta_P) = \mathrm{LF_O}(\theta_O),
\end{aligned}
\label{eq:formu_constrained_theta}
\end{equation}
where $P(\theta_P)$ measures the perceptual error of the perceptual stage output, and $O(\theta_O)$ measures the objective error of the objective stage;
$\mathrm{LF_P}(\theta_P) = T_{\text{LF}}(\mathcal{G}_P(Y^{\prime};\theta_P))$ and $\mathrm{LF_O}(\theta_O) = T_{\text{LF}}(\mathcal{G}_O(X;\theta_O))$ represent the low-frequency subbands of outputs from the perceptual- and objective-focused stage.

The augmented Lagrangian function of \cref{eq:formu_constrained_theta} is:
\begin{equation}
\begin{aligned}
\mathcal{L}(\theta_P,\theta_O,u) =  P(\theta_P) & +O(\theta_O) + u^{T}(\mathrm{LF_P}(\theta_P)\!-\!\mathrm{LF_O}(\theta_O)) \\
    &+ \frac{\rho}{2}\left \| \mathrm{LF_P}(\theta_P) - \mathrm{LF_O}(\theta_O) \right \|_2^2.
\end{aligned}
\end{equation}
where $u$ is the Lagrangian multiplier, and $\rho$ is the penalty parameter. It is often easier to express the above function in the scaled form by defining $u = \rho s$~\cite{zhang2018systematic,kiaee2016alternating,leng2018extremely}, resulting in
\begin{equation}
\begin{aligned}
    \!\!\!\! \mathcal{L}(\theta_P,& \theta_O,s) = P(\theta_P)+O(\theta_O)
    + \frac{\rho}{2}\left \| \mathrm{LF_P}(\theta_P) - \mathrm{LF_O}(\theta_O) + s \right \|^2_2-\frac{\rho}{2}\left \| s \right \|_2^{2}.
    \end{aligned}
    \label{eq:lagrangian_scaled}
\end{equation}
ADMM solves \cref{eq:lagrangian_scaled} by a \emph{decomposition} into three iterative sub-steps:
\begin{subequations}
    \begin{align}
    &\theta_P^{k+1} = \mathop{\argminB}_{\theta_P} \mathcal{L}(\theta_P, \theta_O^{k},s^{k})\label{Ya}\\
    &\theta_O^{k+1} = \mathop{\argminB}_{\theta_O} \mathcal{L}(\theta_P^{k+1}, \theta_O,s^{k})\label{Yb}\\
    &s^{k+1} = s^{k} + \mathrm{LF_P}(\theta_P^{k+1}) - \mathrm{LF_O}(\theta_O^{k+1})\label{Yc}
    \end{align}
    \label{eq:admm_steps}
\end{subequations}
The first two steps are equivalent to:
\begin{subequations}
\begin{align}
&\mathop{\argminB}_{\theta_P} P(\theta_P)+\frac{\rho}{2}\left \| \mathrm{LF_P}(\theta_P)-\mathrm{LF_O}(\theta_O) +s\right \|_2^2 \label{Za}\\
&\mathop{\argminB}_{\theta_O} O(\theta_O)+\frac{\rho}{2}\left \| \mathrm{LF_P}(\theta_P)-\mathrm{LF_O}(\theta_O) +s\right \|_2^2,\label{Zb}
\end{align}
\end{subequations}
where the first term of each minimization function is the loss function of the perception-focused stage or objective-focused stage, \ie $L_P$ or $L_O$ discussed in~\cref{sec:LFc-SR}. The second terms are special $L_2$ regularizers applied to variable $s$ and the low-frequency subbands of the outputs from the two stages. Thus, we can solve these two sub-problems through a deep model optimization algorithm, such as Stochastic Gradient Descent. Based on above analysis through ADMM, our training algorithm for the LFc-SR model consists of three alternating steps and is concluded by dual variable updates in \cref{Yc} after optimizing the perceptual- and objective-focused stage through \cref{Za} and \cref{Zb}. 

\section{Experiments}

\subsection{Settings}
\label{sec:setting}
\textbf{Dataset \& Evaluation:} We used the DIV2K dataset~\cite{timofte2017ntire} for training and validation, We evaluated on standard benchmarks: Set5~\cite{bevilacqua2012low}, Set14~\cite{zeyde2010single}, BSD100~\cite{MartinFTM01}, Urban100~\cite{Huang-CVPR-2015} and Manga109~\cite{mtap_matsui_2017}. To measure objective quality, we computed the PNSR and SSIM on the image's Y channel with Matlab functions. For perceptual quality, we used the no-reference metric NRQM~\cite{ma2017learning} and the full-reference metric LPIPS~\cite{zhang2018unreasonable} for a comprehensive treatment. We provide visual comparisons in ~\cref{fig:visual_compare} and ~\cref{fig:visual_compare_wavelet} to show the qualitative efficacy of our method.

\textbf{Image Decomposition:} We used the discrete wavelet transform (DWT) to decompose our image into frequency subbands. DWT is an invertible frequency transformation common in image processing \cite{zhong2018joint,liu2018multi,guo2017deep}. It decomposes the image into four half-resolution channels, LL, HL, LH and HH, including low- (L) or high- (H) frequency subbands of height and width dimension. DWT can be applied on LL iteratively. We used the LL channel from a 1-level Haar DWT of the image as the LF information in \cref{eq:formu_constrained_theta}. Other transformations are also feasible, and we experimented with Gaussian blur as an ablation. For downsampling function $f_{\downarrow}(\cdot)$ in \cref{eq:stage2}, we also used DWT for convenience; we used the LL channel from a 2-level DWT for $\times 4$ super-resolution. 

\textbf{Model Architecture:} 
We tested our method for $\times4$ super-resolution. For $\mathcal{G}_O$, we directly adopted the architecture of HAN~\cite{niu2020single}, which finish upscaling the images into high resolution. $\mathcal{G}_P$ starts with DWT and ends with inverse DWT (iDWT), saving training time by losslessly reducing the spatial resolution of processed feature maps. Between the two transformations are 15 Res-clique blocks designed by Zhong~\etal~\cite{zhong2018joint}. For the architecture and loss calculation of discriminator for adversarial training, we adopted what was used in SRGAN~\cite{ledig2017photo}. More details are in the Supplementary.

\textbf{Training Details:} We took RGB patches of size $36\!\times\!36$ as inputs and trained the model with the ADAM optimizer~\cite{kingma2014adam} using the settings $\beta_1\!=\!0.9$, $\beta_2\!=\!0.999$, and $\epsilon\!=\!10^{-8}$ on a minibatch of 16. The two stages were pretrained with $L_1$ loss jointly for 400 epochs. The initial learning rate was $10^{-4}$ and decay rates were 0.1 every 100 epochs. We subsequently trained the model using the proposed PD-ADMM algorithm with a penalty factor $\rho\!=\!10^{-4}$ and initialized $s$ with $\mathrm{0}$. We trained the whole model for another 200 epochs with an initial learning rate of $5\times10^{-5}$ and decreasing to one-eighth after 20, 50, 70, 100, and 140 epochs.

\subsection{Comparison with State-of-the-Art Methods}
\label{sec:comparison}
\textbf{Quantitative Comparison with Trade-off Methods:} We first compare our method with existing methods aiming to balance objective and perceptual quality. These include CFSNet~\cite{wang2019cfsnet}, G-MGBP~\cite{navarrete2018multi}, PESR \cite{vu2018perception}, ESRGAN with network interpolation \cite{wang2018esrgan}, Deng \cite{deng2018enhancing}, and WDST \cite{deng2019wavelet}. We note, however, that both Deng and WDST rely on image style transfer~\cite{gatys2016image} for post-processing, which requires much more computational power and impractical inference runtimes (see detailed discussion on inference complexity in \cref{sec:comparison}). Hence, a fair comparison with them is impossible. Nevertheless, as shown in \cref{tab:sota}, our method still reaches comparable performance to WDST and surpasses all other methods.
\begin{table}
\begin{center}
\caption{Comparison with other PD trade-off state-of-the-art models aiming at both objective and perceptual quality on $\times 4$ super-resolution. We color the best performance in \textcolor{red}{red} and second best in \textcolor{blue}{blue}. We separate Deng~\cite{deng2018enhancing} and WDST~\cite{deng2019wavelet} as they rely on three orders of magnitude higher computational power and a fair comparison with other listed methods is impossible. For reference, we set the best performance of Deng and WDST that surpass all other methods to \textbf{bold}. Higher PSNR, SSIM, NRQM score and \emph{lower} LPIPS score mean better performance.}
\label{tab:sota}
\resizebox{\textwidth}{!}{
\begin{tabular}{ c | c|c c| c c c c c  }
 \hline
 \hline
  Dateset&Metric & Deng~\cite{deng2018enhancing}&WDST~\cite{deng2019wavelet}& G-MGBP~\cite{navarrete2018multi} & PESR~\cite{vu2018perception}  & ESRGAN~\cite{wang2018esrgan}  &CFSNet~\cite{wang2019cfsnet}&Ours\\
 \hline
 \multirow{4}{*}{Set5~\cite{bevilacqua2012low}}&PSNR & 31.14&31.46&30.87 & 30.76& \color{blue}{31.11}& 31.00&\color{red}{31.79}\\
 &SSIM & 0.8917&\textbf{0.8929}&0.8807 & \color{red}{0.8915} &0.8839&0.8894 & \color{blue}{0.8910}\\
 &NRQM &7.0022&\textbf{7.5180}&7.3155&7.1344  &7.0724&\color{red}{7.4820}&\color{blue}{7.3462}\\
 &LPIPS &-& 0.0868&0.1003&0.0884 &\color{blue}{0.0841}& 0.1020&\color{red}{0.0776} \\
 \hline
 \multirow{4}{*}{Set14~\cite{kim2010single}}&PSNR &27.77&\textbf{28.07}&27.56 &27.57 &27.53& \color{blue}{27.61}&\color{red}{27.87}\\
 &SSIM & 0.8325&\textbf{0.8356}&0.8206 &\color{red}{0.8322}  &0.8228&0.8280  & \color{blue}{0.8282} \\
 &NRQM &7.5575 &7.6827 &7.5042 &7.5301 &7.5936&\color{blue}{7.6074} &\color{red}{7.7957}\\
 &LPIPS& -& 0.1658 &0.1757 &0.1612 &\color{blue}{0.1539}&0.1754 &\color{red}{0.1423} \\
 \hline
 \multirow{4}{*}{BSD100~\cite{MartinFTM01}}&PSNR &26.46&26.82&\color{blue}{26.59} &26.33  &26.44&26.46 &\color{red}{26.84}\\
 &SSIM &0.7048&\textbf{0.7085}&0.6926 & 0.6980 &\color{blue}{0.7002}&0.6991 &\color{red}{0.7010} \\
 &NRQM &8.4452&\textbf{8.5948}&8.1790 &8.3298 &8.3034&\color{blue}{8.3770} &\color{red}{8.4406}\\
 &LPIPS & -&0.2140&0.2238 &0.2045 &\color{blue}{0.1903} &0.2129 &\color{red}{0.187} \\
 \hline
 \multirow{4}{*}{Urban100~\cite{Huang-CVPR-2015}}&PSNR &25.96&26.26&25.15 &25.88  &\color{blue}{26.08}&25.38 &\color{red}{26.28}\\
 &SSIM & 0.9620&\textbf{0.9649}&0.9495 &0.9610 & \color{blue}{0.9624}&0.9546 & \color{red}{0.9636} \\
 &NRQM &6.4317&6.4556&6.2190 &6.3190 &6.1762&\color{blue}{6.5140}&\color{red}{6.6220}\\
 &LPIPS & -& 0.1604&0.1775 &\color{blue}{0.1402} & 0.1519&0.1506 & \color{red}{0.1235}\\
 \hline
  \multirow{4}{*}{Manga109~\cite{mtap_matsui_2017}}&PSNR & -&-& 29.07 &28.77& \color{blue}{29.72} &29.49 & \color{red}{30.20}\\
 &SSIM &-&-& 0.8815 & \color{blue}{0.9795} &0.9772&0.9789 & \color{red}{0.9804}\\
 &NRQM &-&-&6.4073 & \color{blue}{6.6071}  &6.3909 &6.5344 & \color{red}{6.6532} \\
 &LPIPS &-&-&0.0779 & 0.0634 & \color{red}{0.0610} &0.0719 & \color{blue}{0.0627} \\
 \hline
\end{tabular}}
\end{center}
\end{table}

\begin{figure}[h!]
    \centering
    \begin{subfigure}[t]{0.49\textwidth}
        \includegraphics[width=0.9\linewidth]{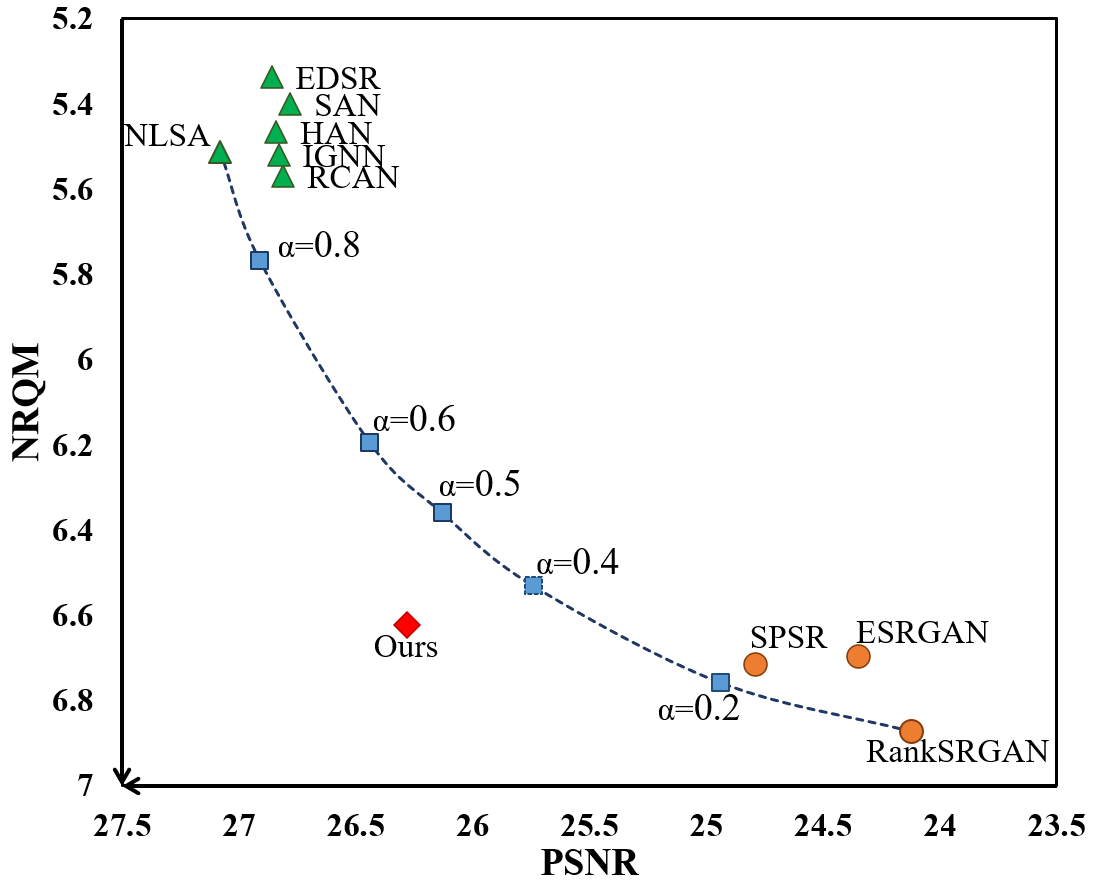}
        \subcaption{Performance on Urban100~\cite{Huang-CVPR-2015}} 
    \end{subfigure}
    \begin{subfigure}[t]{0.49\textwidth}
        \includegraphics[width=0.9\linewidth]{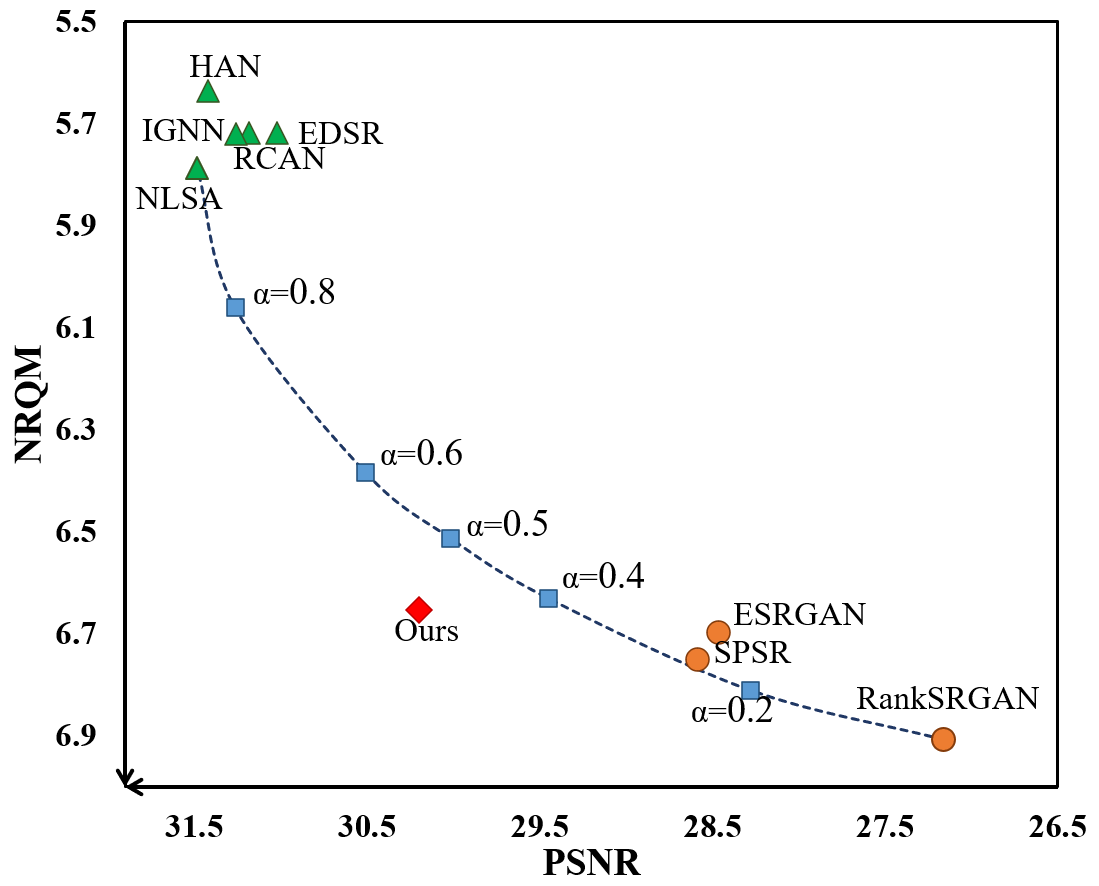}
        \subcaption{Performance on Manga109~\cite{mtap_matsui_2017}}
    \end{subfigure}
    \caption{Comparison with single-focused start-of-the-art SR methods on Urban100~\cite{Huang-CVPR-2015} and Manga109~\cite{mtap_matsui_2017} datasets for $\times4$ super-resolution. We plot the curve by interpolating results from the best objective estimator~\cite{mei2021image} and the best perceptual estimator~\cite{zhang2019ranksrgan} on the plane. Our model can reach a better balance than the trade-off boundary approximated by single-focused state-of-the-art methods, as it sits well below the boundary.}
    \label{fig:sota_single}
\end{figure}

\textbf{Quantitative Comparison with Single-focused Methods:}
We further compare our method with single-focused state-of-the-art methods. Quantitatively comparing our balanced method with methods that only consider one type of quality is non-trivial. Due to the PD trade-off, our method will be inferior to the single-focused method in its optimized direction while surpassing it in the other. We thus make the comparison on a PD trade-off plane~\cite{blau2018perception}. We generate a curve on the plane by interpolating the outputs from the state-of-the-art objective-focused and perceptual-focused methods. The interpolated image $Y$ is produced through the following function:
\begin{equation}
    Y = \alpha \cdot Y^O + (1-\alpha) \cdot Y^P,
    \label{eq:curve}
\end{equation}
where $Y^O$ is an HR image super-resolved by objective-focused model, $Y^P$ is the perceptual-focused counterpart and the value of $\alpha$ is in $[0,1]$. 
Freirich~\etal~\cite{freirich2021theory} found this interpolation curve to be an optimal PD trade-off boundary achievable using single-focused estimators.
Methods that sit in the region below the curve are better than those on or above the curve in terms of PD trade-off. In~\cref{fig:sota_single}, we plot the curve between the best estimators, NLSA~\cite{mei2021image} and RankSRGAN~\cite{zhang2019ranksrgan}, and show that our method is the only one below the estimated trade-off boundary considering all single-focused state-of-the-art methods on the plane.

\begin{figure}[t]
    \centering
    \begin{subfigure}[t]{0.19\textwidth}
    \centering
        \includegraphics[width=\linewidth]{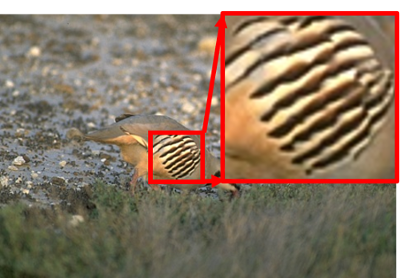}
    \end{subfigure}
    \begin{subfigure}[t]{0.19\textwidth}
    \centering
        \includegraphics[width=\linewidth]{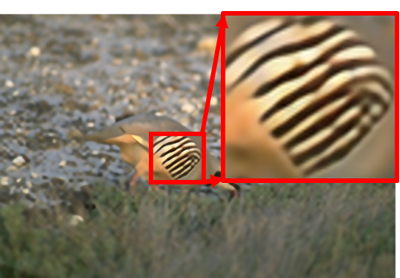}
    \end{subfigure}
    \begin{subfigure}[t]{0.19\textwidth}
    \centering
        \includegraphics[width=\linewidth]{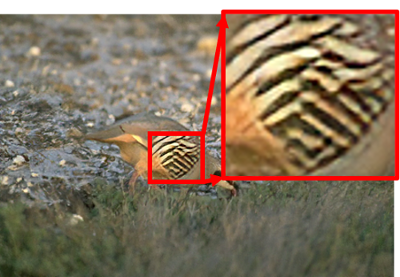}
    \end{subfigure}
    \begin{subfigure}[t]{0.19\textwidth}
    \centering
        \includegraphics[width=\linewidth]{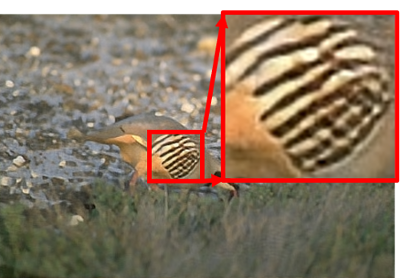}
    \end{subfigure}
    \begin{subfigure}[t]{0.19\textwidth}
    \centering
        \includegraphics[width=\linewidth]{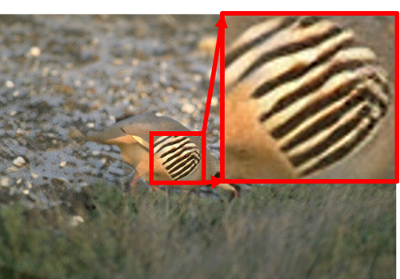}
    \end{subfigure}
    \begin{subfigure}[t]{0.19\textwidth}
    \centering
        \includegraphics[width=\linewidth]{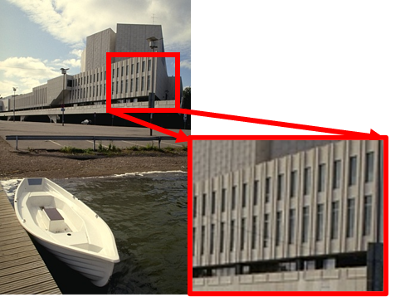}
    \end{subfigure}
    \begin{subfigure}[t]{0.19\textwidth}
    \centering
        \includegraphics[width=\linewidth]{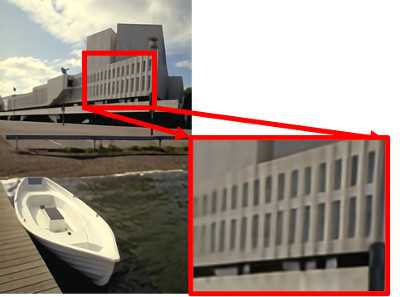}
    \end{subfigure}
    \begin{subfigure}[t]{0.19\textwidth}
    \centering
        \includegraphics[width=\linewidth]{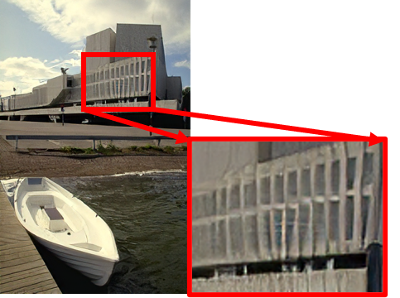}
    \end{subfigure}
    \begin{subfigure}[t]{0.19\textwidth}
    \centering
        \includegraphics[width=\linewidth]{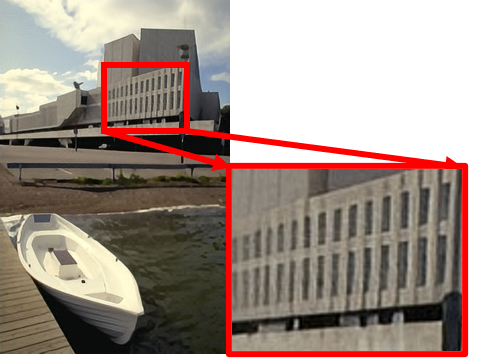}
    \end{subfigure}
    \begin{subfigure}[t]{0.19\textwidth}
    \centering
        \includegraphics[width=\linewidth]{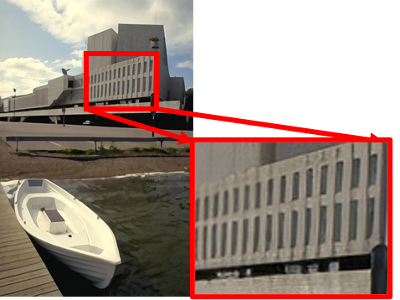}
    \end{subfigure}
    \begin{subfigure}[t]{0.19\textwidth}
    \centering
        \includegraphics[width=\linewidth]{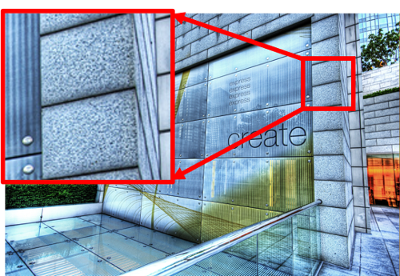}
    \end{subfigure}
    \begin{subfigure}[t]{0.19\textwidth}
    \centering
        \includegraphics[width=\linewidth]{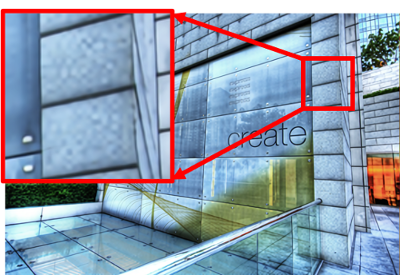}
    \end{subfigure}
    \begin{subfigure}[t]{0.19\textwidth}
    \centering
        \includegraphics[width=\linewidth]{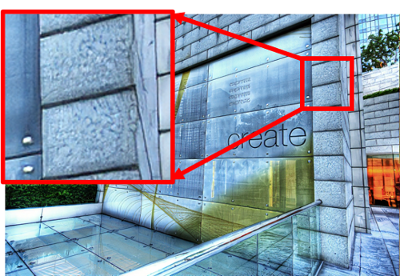}
    \end{subfigure}
    \begin{subfigure}[t]{0.19\textwidth}
    \centering
        \includegraphics[width=\linewidth]{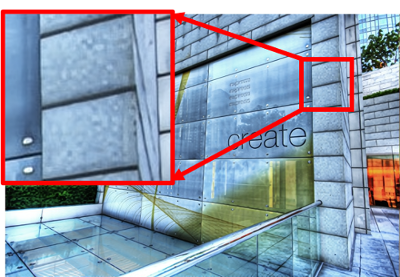}
    \end{subfigure}
    \begin{subfigure}[t]{0.19\textwidth}
    \centering
        \includegraphics[width=\linewidth]{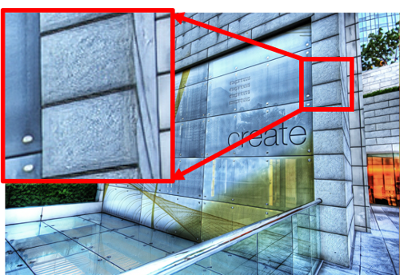}
    \end{subfigure}\\
    \begin{subfigure}[t]{0.19\textwidth}
    \centering
        \includegraphics[width=\linewidth]{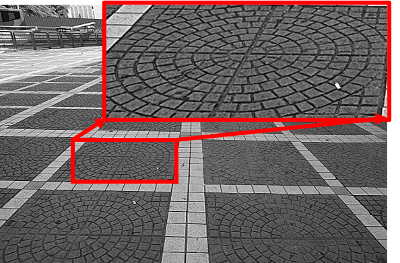}
        \subcaption{HR}
    \end{subfigure}
    \begin{subfigure}[t]{0.19\textwidth}
        \centering
        \includegraphics[width=\linewidth]{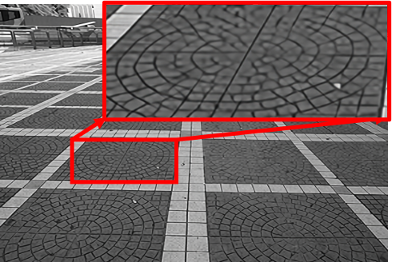}
        \subcaption{NLSA}
    \end{subfigure}
    \begin{subfigure}[t]{0.19\textwidth}
        \centering
        \includegraphics[width=\linewidth]{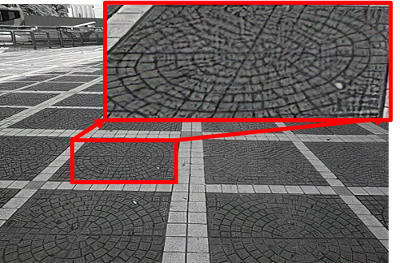}
        \subcaption{Ranksrgan}
    \end{subfigure}
    \begin{subfigure}[t]{0.19\textwidth}
    \centering
        \includegraphics[width=\linewidth]{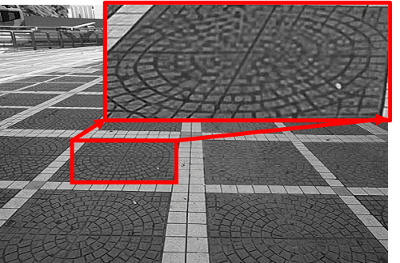}
        \subcaption{WDST}
    \end{subfigure}
    \begin{subfigure}[t]{0.19\textwidth}
    \centering
        \includegraphics[width=\linewidth]{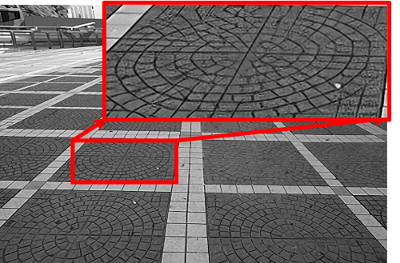}
        \subcaption{Our Result}
    \end{subfigure}
    \caption{Image sample comparison for $\times4$ super-resolution on BSD100~\cite{MartinFTM01} \textit{(first two rows)} and Urban100 \cite{Huang-CVPR-2015} \textit{(thrid and forth rows)} dataset. NLSA~\cite{mei2021image} is objective-focused, RankSRGAN~\cite{zhang2019ranksrgan} is perceptual-focused, while WDST~\cite{deng2019wavelet} and our method are balanced approaches. Our method retains relatively sharp detailing without introducing unnatural artifacts.}
    \label{fig:visual_compare}
\end{figure}

\begin{figure}[h!]
    \centering
    \begin{subfigure}[t]{0.24\textwidth}
    \centering
        \includegraphics[width=0.9\linewidth]{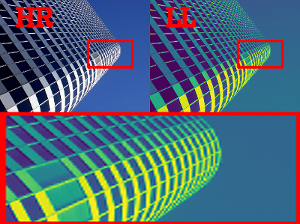}
    \end{subfigure}    
    \begin{subfigure}[t]{0.24\textwidth}
    \centering
        \includegraphics[width=0.9\linewidth]{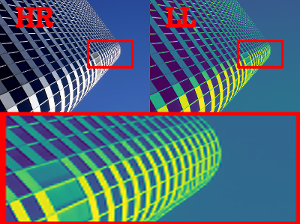}
    \end{subfigure}
    \begin{subfigure}[t]{0.24\textwidth}
    \centering
        \includegraphics[width=0.9\linewidth]{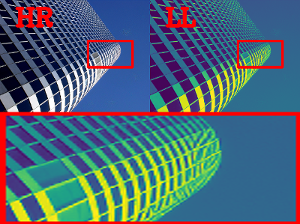}
    \end{subfigure}
    \begin{subfigure}[t]{0.24\textwidth}
    \centering
        \includegraphics[width=0.9\linewidth]{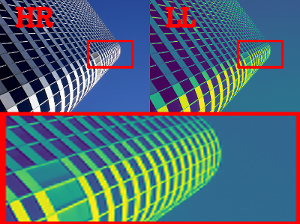}
    \end{subfigure}\\
    \begin{subfigure}[t]{0.24\textwidth}
    \centering
        \includegraphics[width=0.9\linewidth,height=0.7\linewidth]{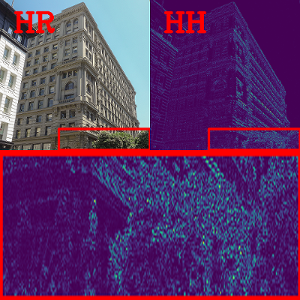}
        \subcaption{Ground Truth}
    \end{subfigure}
    \begin{subfigure}[t]{0.24\textwidth}
    \centering
        \includegraphics[width=0.9\linewidth, height=0.7\linewidth]{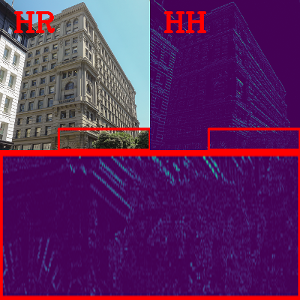}
        \subcaption{NLSA}
    \end{subfigure}
    \begin{subfigure}[t]{0.24\textwidth}
    \centering
        \includegraphics[width=0.9\linewidth,height=0.7\linewidth]{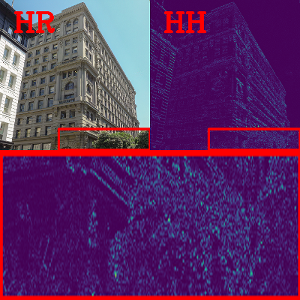}
        \subcaption{RankSRGAN}
    \end{subfigure}
    \begin{subfigure}[t]{0.24\textwidth}
    \centering
        \includegraphics[width=0.9\linewidth,height=0.7\linewidth]{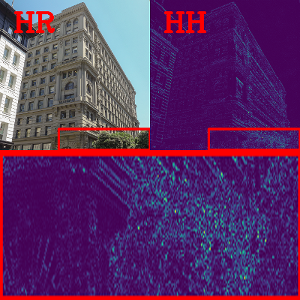}
        \subcaption{Our Result}
    \end{subfigure}
    \caption{Visual comparison of low- and high-frequency subbands for $\times4$ super-resolution by NLSA~\cite{mei2021image}, RankSRGAN~\cite{zhang2019ranksrgan} and our method. We visualize the LL \textit{(first row)} and HH \textit{(second row)} wavelet channels for low- and high-frequency comparison. \emph{Within each image}, we show the HR image \textit{(top left)}, wavelet channel \textit{(top right)} and the zoom-in region of the wavelet channel under them. NLSA generates high-fidelity low-frequency information but lacks high-frequency detailing in the HH channel; RankSRGAN produces twisted structures in the LL channel but can reconstruct more high-frequency information. Our method predicts both high-fidelity LL channel and abundant high-frequency detailing.
    }
    \label{fig:visual_compare_wavelet}
\end{figure}

 \textbf{Qualitative Results:} We compare image samples generated from our method with those from NLSA \cite{mei2021image}, RankSRGAN \cite{zhang2019ranksrgan} and WDST \cite{deng2019wavelet} on BSD100 \cite{MartinFTM01} and Urban100 \cite{Huang-CVPR-2015} in \cref{fig:visual_compare}. Our method generates clear structures with fewer artifacts compared to single-focused methods and more realistic high-frequency details than WDST does. We provide frequency subbands comparison in \cref{fig:visual_compare_wavelet}. While single-focused methods only achieve good results at one of the subbands, our method learns both low- and high-frequency information effectively.

\textbf{Inference Complexity:}
A significant advantage of our method is that it is one-shot. We directly optimize a single model and bypass the complex fusing of two outputs, unlike competing work WDST~\cite{deng2019wavelet} and its predecessor, Deng~\cite{deng2018enhancing}. Central to their inference is an image style transfer procedure~\cite{gatys2016image}, which requires thousands of iterations to update the initial input signal towards a fused wavelet channel or image. In contrast, our method super-resolves the LR image through a single model in a one-shot manner. We used a 24GB NVIDIA RTX A5000 GPU for running the models. \cref{tab:inference_complexity} provides a quantitative comparison of the computational complexity, including the FLOPs and runtime for $\times 4$ super-resolution of a $128 \times 128$ LR input. Our model's FLOPs and runtime are three orders of magnitude lower than WDST, assuming only 1000 iterations of updates for WDST (the actual number of updates is usually much more). 

\begin{table}
\begin{center}
\caption{Comparison of the inference complexity of our method and WDST~\cite{deng2019wavelet} for $\times4$ super-resolution of a $128 \times 128$ LR image. The SR model super-resolves LR inputs and gives HR images, while post-processing further process super-resolved images. Our model's FLOPs and runtime are three orders of magnitude less than WDST. We refer to the supplementary material for detailed analysis.}
\label{tab:inference_complexity}
\begin{tabular}{c|c|c c |c}
\hline
\hline
Method & Values & SR model & Post-processing & All\\
\hline
\multirow{2}{*}{WDST}&FLOPs&50.6M &$\approx 37448.1M$ & $\approx 37500.3M$\\
&Run-time & 0.918s& $\approx 1152s$&$\approx 1153s$ \\
\hline
\multirow{2}{*}{Ours}&FLOPs&26.8M &- &26.8M\\
 &Run-time & 0.566s & -&0.566s \\
\hline
\end{tabular}
\end{center}
\end{table}

\subsection{Ablation Studies}
\begin{figure}[h!]
    \centering
    \begin{subfigure}[t]{0.48\textwidth}
        \centering
        \includegraphics[width=\linewidth,height=0.85\linewidth]{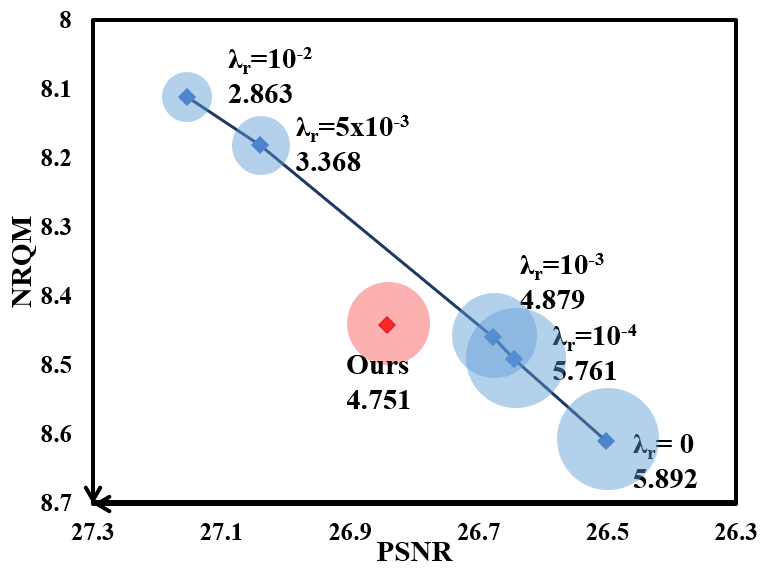}
        \subcaption{Performance on BSD100~\cite{MartinFTM01}} 
    \end{subfigure}
    \begin{subfigure}[t]{0.48\textwidth}
        \centering
        \includegraphics[width=\linewidth,height=0.85\linewidth]{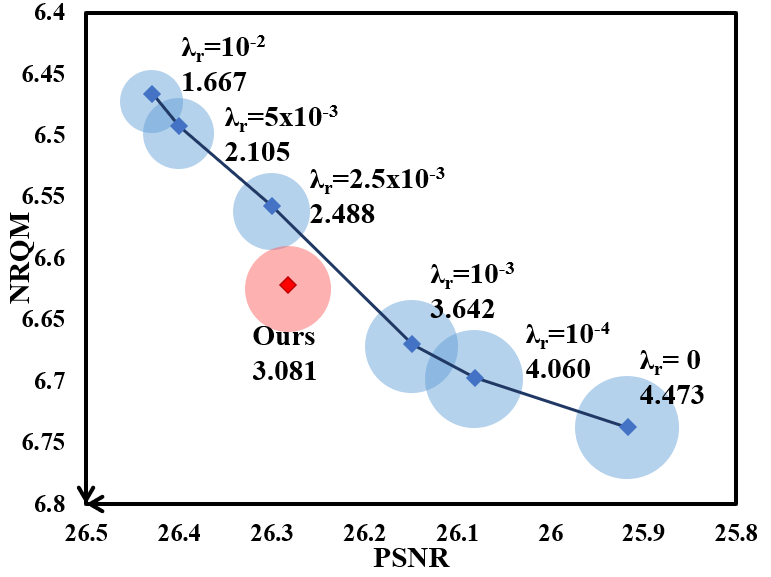}
        \subcaption{Performance on Urban100~\cite{Huang-CVPR-2015}}
    \end{subfigure}
    \caption{Comparison with regularizer-based method on BSD100~\cite{MartinFTM01} and Urban100~\cite{Huang-CVPR-2015}. $\lambda_r$ is the weight of regularizer; numbers below $\lambda_r$ are mean absolute error (MAE) of low-frequency subbands between outputs of two stages. A lower value means a higher similarity. Our method achieves appropriate LF similarity and better balance than the curve plot by regularizer-based models.
    }
    \label{fig:abla_reg}
\end{figure}

\begin{figure}[h!]
    \centering
    \begin{subfigure}[t]{0.24\textwidth}
    \centering
    \includegraphics[width=0.85\linewidth]{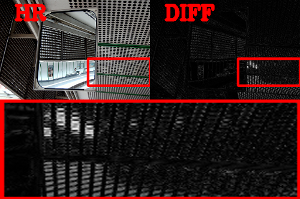}
    \subcaption{w/o constraint}
    \end{subfigure}
    \begin{subfigure}[t]{0.24\textwidth}
    \centering
        \includegraphics[width=0.85\linewidth]{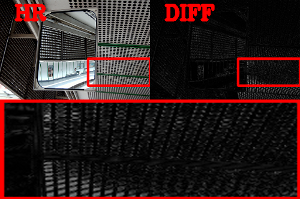}
        \subcaption{PD-ADMM}
    \end{subfigure}
    \begin{subfigure}[t]{0.24\textwidth}
    \centering
        \includegraphics[width=0.85\linewidth]{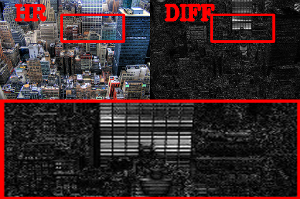}
        \subcaption{w/o constraint}
    \end{subfigure}
    \begin{subfigure}[t]{0.24\textwidth}
    \centering
        \includegraphics[width=0.85\linewidth]{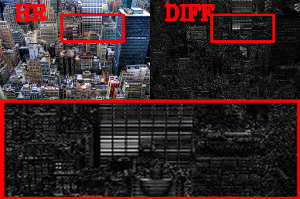}
        \subcaption{PD-ADMM}
    \end{subfigure}
    \caption{Visualization of \emph{difference} in low-frequency subband compared to the ground-truth for $\times 4$ super-resolution. Each image shows the HR result \textit{(top left)}, grayscale map of difference of the LL channel between the variant and the ground truth \textit{(top right)}, and the zoom-in region of the grayscale map. The darker the map, the better. Limiting the change of the LF subband keeps higher accuracy of the LF information while learning the HF information at the perceptual stage.}
    \label{fig:visual_compare_woconstraint}
\end{figure}

\textbf{Verification of low-frequency constraint and ADMM optimization:}
In this experiment, we kept a constant architecture as described in \cref{sec:setting}. Then we implemented the equivalence constraint through the second-stage regularizer in \cref{eq:ll_reg} and trained the models with different weights $\lambda_r$ for the regularizer.~\cref{fig:abla_reg} shows the performance of regularizer-based methods on the PD trade-off plane, together with our PD-ADMM method; for each hyperparameter setting, we also illustrate the LF difference between the two stages' outputs, \ie $Y^{\prime}$ and $Y$, with a circle. The more different the two stages, the larger the circle.
Compared to the no-constraint version ($\lambda_r=0$), both using PD-ADMM and adding the low-frequency constraint yield higher LF similarity. \cref{fig:visual_compare_woconstraint} also provides visual evidence of the efficacy of adding the low-frequency constraint. The increase of $\lambda_r$ results in higher LF similarity, transitioning from better perceptual quality to better objective quality. However, the model optimized by PD-ADMM achieves a superior balance of perceptual- and objective-quality to this transition.

\begin{figure}[h!]
    \centering
    \begin{subfigure}[t]{0.48\textwidth}   
    \centering
        \includegraphics[width=0.95\linewidth,height=0.90\linewidth]{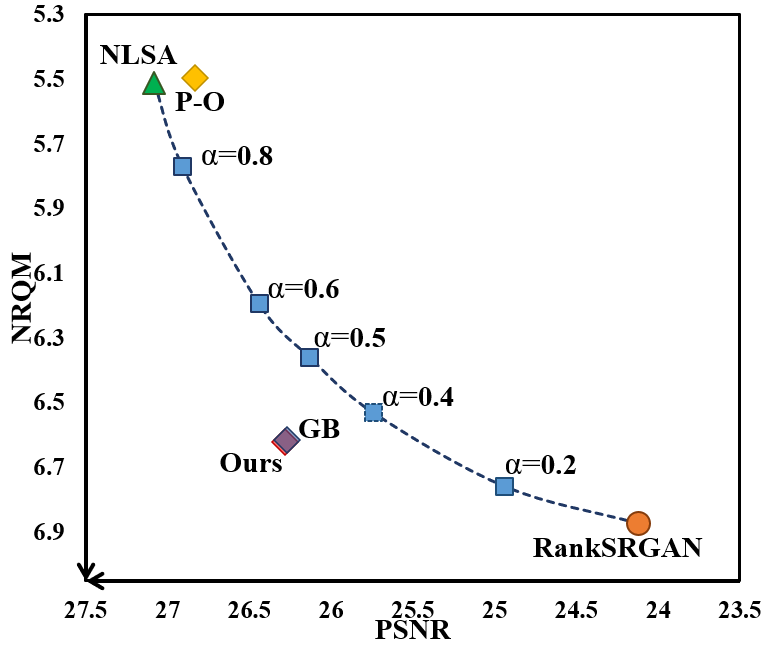}
        \subcaption{Urban100\cite{Huang-CVPR-2015}}
    \end{subfigure}
    \begin{subfigure}[t]{0.48\textwidth}
    \centering
        \includegraphics[width=0.95\linewidth,height=0.90\linewidth]{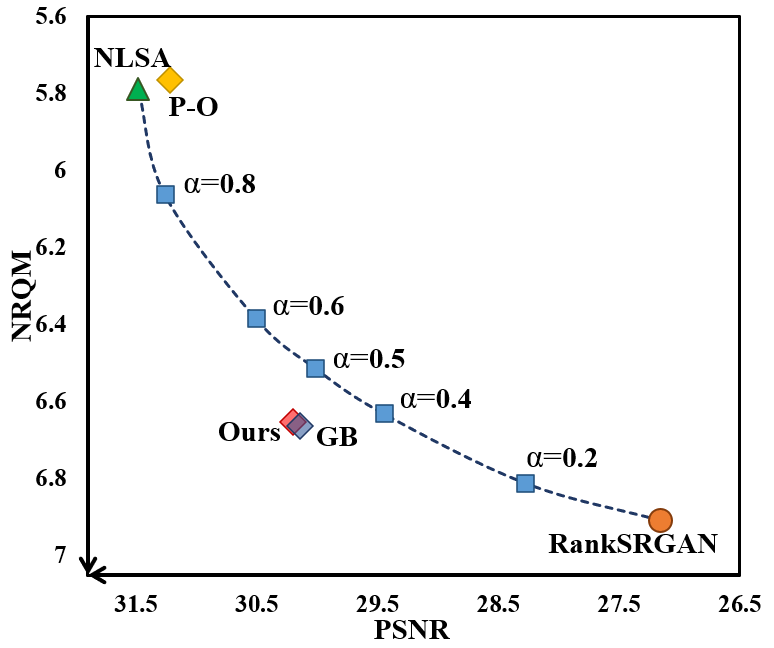}
        \subcaption{Manga109\cite{mtap_matsui_2017}}
    \end{subfigure}
    \caption{Comparison between our model and its two variants. The order-exchanged version \textit{(P-O)} optimizes HF subband for perceptual quality first. Its performance is close to the top-left corner and above the estimated trade-off boundary, \ie it fails to balance the perceptual and objective quality. The other variant \textit{(GB)} implements the low-frequency constraint of LFc-SR by Gaussian Blur rather than DWT used in our original model. The close performance of two models shows that our constraint design does not rely on a specific extraction method.}
    \label{fig:ablation_2_3}
\end{figure}

\textbf{Stage Order and Constrained Subband:} This experiment validates the reconstruction ordering of frequency subbands in our proposed model. Without changing model architecture, we exchanged the optimization objectives of the two stages and converted the equivalence constraint to HF subbands extracted by DWT. Representing our original method as the O-P model, this swapped P-O model completes perceptual-focused reconstruction first, and then optimizes LF subbands while keeping HF equivalence to the first stage. We also optimized the P-O model through PD-ADMM. Borrowing the comparison with single-focused models in \cref{sec:comparison}, we plotted the results of our original model and P-O model on a PD trade-off plane with the trade-off boundary. As shown in \cref{fig:ablation_2_3}, the performance of the order-exchanged model is dominated by the objective optimization and fails to achieve a good PD trade-off.

\textbf{Low-frequency Subbands Extraction:} 
To verify that our results are not dependent on specific frequency decomposition method, \ie the DWT, we applied Gaussian blur on the outputs from the two stages to extract the LF subbands that should satisfy the equivalence constraint in PD-ADMM. 
Experimentally, we applied convolution with a $21\times 21$ Gaussian kernel with $\sigma = 3$ to extract the LF information. As shown in \cref{fig:ablation_2_3}, models with DWT and with Gaussian Blur (GB) have extremely close positions on the PD trade-off plane, and both of them are under the PD trade-off boundary introduced in \cref{sec:comparison}. It shows that using different extraction methods can reach similar PD balances. We chose DWT for our model because of its popularity in image frequency analysis.
\section{Conclusion \& Limitations}
In this paper, we learned a SISR model for both objective and perceptual quality. We proposed LFc-SR, a two-stage model trained with a low-frequency constraint to implicitly optimize low- and high-frequency bands for objective and perceptual quality in a successive way. The training of LFc-SR is formulated as a constrained multi-objective optimization problem. We designed PD-ADMM, an alternating algorithm, to solve for a solution that balances the perception-distortion trade-off. Our method is effective compared with other PD trade-off related methods and single-focused models without the expense of heavy fusion procedure.

Despite being more efficient than trade-off related fusion methods, our method is still far from optimal compared with some efficiency-oriented SISR models. Due to the successive processing in stages, some parts of the two-stage model may extract similar information and cause redundancy. However, such a design also offers the potential to reuse information efficiently, which we will leave for future work.

\subsubsection{Acknowledgement}
This research is supported by Singapore Ministry of Education (MOE) Academic Research Fund Tier 1 T1251RES1819.
\clearpage
%
%
\bibliographystyle{splncs04}
\bibliography{egbib}
\end{document}